\documentclass[twocolumn,aps]{revtex4}

\newcommand{\be}{\begin{equation}}
\newcommand{\ee}{\end{equation}}
\newcommand{\beq}{\begin{eqnarray}}
\newcommand{\eeq}{\end{eqnarray}}
\newcommand{\noi}{\noindent}
\newcommand{\gsim}{\stackrel{>}{\sim}}

\begin{document}

\title{\large\bf Precise Determination of $|V_{us}|$ from Lattice Calculations of \\
Pseudoscalar Decay Constants}

\author{William J. Marciano}
\altaffiliation{Work supported by DOE Grant DE-AC02-98CH10886} 

\affiliation{Brookhaven National Laboratory \\ Upton, New York\ \ 11973}

\date{\today}

\begin{abstract}
Combining the ratio of experimental kaon and pion decay widths, $\Gamma(K\to \mu\bar\nu_\mu(\gamma))/\Gamma(\pi\to\mu \bar\nu_\mu(\gamma))$, with a recent lattice gauge theory calculation of $f_K/f_\pi$ provides a precise value for the CKM quark mixing matrix element $|V_{us}|=0.2236(30)$ or if 3 generation unitarity is assumed $|V_{us}|=0.2238(30)$. Comparison with other determinations of that fundamental parameter, implications, and an outlook for future improvements are given.
\end{abstract}

\maketitle

Recently \cite{davies}, high precision lattice QCD results have been obtained for a number of interesting phenomenological quantities. Those first principles theory calculations already provide impressive confrontations with experiments at the $\pm3\%$ (or better) level and further improvement is expected as computer power increases and new lattice techniques are applied.

The pion and kaon decay constants, $f_\pi$ and $f_K$, are among the newly lattice calculated quantities. Preliminary values have been obtained \cite{aubin}

\beq
f_\pi & = & 129.3 \pm 1.1 \pm 3.5{\rm ~MeV} \label{eqone} \\
f_K & = & 155.0 \pm 1.8 \pm 3.7 {\rm ~MeV} \label{eqtwo} \\
f_K/f_\pi & = & 1.201 (8) (15) \label{eqthree}
\eeq

\noi where the first error is statistical and the second systematic. The scale uncertainty of $\pm2.2\%$ dominates $f_\pi$ and $f_K$ individually, but largely cancels in the ratio. For that reason, $f_K/f_\pi$ has a smaller relative systematic error of only $\pm1.2\%$, stemming largely from chiral and continuum extrapolations \cite{gott}. In addition, the statistical errors are correlated and partially cancel in the ratio. Although the specific numbers in eqs.~(\ref{eqone})--(\ref{eqthree}) are labeled as preliminary, the ratio $f_K/f_\pi$ should be rather stable because of its scale determination insensitivity.

In this letter, I point out that the result for $f_K/f_\pi$ in eq.~(\ref{eqthree}) can be used to provide a very accurate determination of the CKM quark mixing matrix element $|V_{us}|$. In fact, the procedure I describe is already competitive with other, more traditional measurements of that important parameter. Its addition to those other approaches is particularly welcome because a long standing controversy exists regarding the actual value of $|V_{us}|$. The generally accepted PDG \cite{hagi,renk} value

\be
|V_{us}| = 0.2196 (26) \qquad {\rm PDG~~2002} \label{eqfour}
\ee

\noi based on an average of relatively old $K_{e3} (K\to\pi e\nu$) decay rates,  combined with the value of $|V_{ud}|$ obtained from super-allowed $0^+\to0^+$ nuclear beta decays \cite{renk}

\be
|V_{ud}| = 0.9740 \pm 0.0005 \label{eqfive}
\ee

\noi and the fact that $|V_{ub}|$ is negligible gives

\be
|V_{ud}|^2 + |V_{us}|^2 +|V_{ub}|^2 = 0.9969 (15) \label{eqsix}
\ee

\noi a 2 sigma deviation from the 3 Generation CKM unitarity expectation of 1. However, recent studies of $K_{e3}$ \cite{sher} and Hyperon decays \cite{cabib} suggest larger values for $|V_{us}|$

\beq
& |V_{us}| = 0.2272 (30) & {\rm E865~} K_{e3} [6]  \label{eqseven} \\
& |V_{us}| = 0.2250 (27) & {\rm Hyperon~Decays} [7] \label{eqeight} 
\eeq

\noi consistent with unitarity. Resolution of this discrepancy is an outstanding problem for particle and nuclear physics.

The actual value of $|V_{us}|$ is also important for other reasons. It provides the $\lambda=|V_{us}|$ parameter (also known as $\sin\theta_{\rm Cabibbo}$) of the Wolfenstein \cite{wolf} CKM matrix parametrization. In fact, $\lambda$ is the cornerstone of that formalism and as such influences values of its other 3 parameters ($A, \rho, \eta$) as well as the Standard Model predictions for CP violation and rare decay rates \cite{marcone}. So, determining $|V_{us}|$ as precisely and dependably as possible is critically important. In that regard, the approach advocated  here, employing the lattice $f_K/f_\pi$ value as input, not only provides a currently competitive determination of $|V_{us}|$, but offers the possibility for further significant improvement in the future.

My starting point is the calculated decay rates for $\pi\to\mu\bar\nu_\mu (\gamma)$ and $K\to\mu\bar\nu_\mu (\gamma)$ (called $\pi_{\mu2}$ and $K_{\mu2}$ in the literature) which are usually used to extract $f_\pi$ and $f_K$. Here $(\gamma)$ indicates that radiative inclusive decay rates $\mu\bar\nu_\mu + \mu\bar\nu_\mu \gamma + \mu\bar\nu \gamma\gamma\dots$ are implied. Those decay rates are given by \cite{marctwo}.

\begin{widetext}
\beq
\Gamma(\pi\to\mu\bar\nu_\mu(\gamma)) & = & \frac{G^2_\mu |V_{ud}|^2}{8\pi} f^2_\pi m_\pi m^2_\mu \left( 1-\frac{m^2_\mu}{m^2_\pi} \right)^2 \left[1+ \frac{\alpha}{\pi} C_\pi\right] \label{eqnine} \\
\Gamma(K\to\mu\bar\nu_\mu(\gamma)) & = & \frac{G^2_\mu |V_{us}|^2}{8\pi} f^2_K m_K m^2_\mu \left( 1-\frac{m^2_\mu}{m^2_K} \right)^2 \left[1+ \frac{\alpha}{\pi} C_K\right] \label{eqten}
\eeq
\end{widetext}
\noi where

\beq
G_\mu & = & 1.16637 (1) \times 10^{-5} {\rm ~GeV}^{-2} \nonumber \\
m_\mu & = & 105.658357 {\rm ~MeV} \nonumber \\
m_\pi & = & 139.57018 (35) {\rm ~MeV} \label{eqeleven} \\
m_K & = & 493.677 (13) {\rm ~MeV} \nonumber
\eeq

\noi are very precisely known \cite{hagi}. The radiative inclusive electroweak corrections in those expressions are parametrized by $C_\pi$ and $C_K$, They include virtual loop effects as well as real bremsstrahlung emission. By convention, all ${\cal{O}}(\alpha)$ effects have been factored out in eqs.~(\ref{eqnine}) and (\ref{eqten}); so, $f_\pi$ and $f_K$ should in principle be independent of QED ambiguities \cite{hagi,marctwo}. 

The largest radiative corrections in eqs.~(\ref{eqnine}) and (\ref{eqten}) are $+2.4\%$ short-distance effects which are the same (universal) for $C_\pi$ and $C_K$ \cite{sirlin,marcthree}. A second important contribution is the long distance radiative correction appropriate for point-like (elementary) pions and kaons calculated long ago by Kinoshita \cite{kino}. That contribution called \cite{marctwo} $F(x), x=m_\mu/m_\pi$ or $m_\mu/m_K$ gives rise to a 0.19\% difference in $\pi$ and $K$ decays with essentially no uncertainty. The only real uncertainty in $C_\pi$ and $C_K$ stems from hadronic structure dependent radiative corrections, virtual and bremsstrahlung \cite{marcfour}. (Although structure dependent bremsstrahlung effects are very small.) Those corrections must be computed in a model of hadronic structure \cite{fink}, usually parametrized by form factors required to properly \cite{marcfive} extrapolate between long and short-distance effects. An estimate of those structure dependent corrections by Finkemeier \cite{fink} leads to an overall difference

\be
C_\pi - C_K = 3.0 \pm 1.5 \label{eqtwelve}
\ee

\noi where a rather generous error, due to structure dependence has been assigned. However, even increasing the error in eq.~(\ref{eqtwelve}) by a factor of 2 would have little effect on the results in this paper or possible future improvements. A detailed expose on the hadronic structure dependent corrections and their uncertainties will be presented in a subsequent publication which will critique the study in ref. \cite{fink}.

Using the result in eq.~(\ref{eqtwelve}), one finds for the ratio of decay rates

\beq
\frac{\Gamma(K\to\mu \bar\nu_\mu(\gamma))}{\Gamma(\pi\to\mu\bar\nu_\mu (\gamma))} & = & \frac{|V_{us}|^2 f^2_K m_K \left(1-\frac{m^2_\mu}{m^2_K}\right)^2} {|V_{ud}|^2 f^2_\pi m_\pi \left(1-\frac{m^2_\mu}{m^2_\pi}\right)^2} \nonumber \\
 & & \times (0.9930(35)) \label{eqthirteen} 
\eeq

\noi Next, employing the experimental values \cite{hagi}

\be
\begin{array}{rl}
\Gamma(\pi\to\mu\bar\nu_\mu (\gamma)) & =  2.528 (2)\times10^{-14} {\rm ~MeV} \\
\Gamma(K\to\mu\bar\nu_\mu (\gamma)) & = 3.372 (9)\times10^{-14} {\rm ~MeV} 
\end{array}
\label{eqfourteen}
\ee

\noi gives

\be
\frac{\Gamma(K\to\mu\bar\nu_\mu (\gamma))}{\Gamma (\pi\to\mu\bar\nu_\mu (\gamma))} = 1.334 (4) \label{eqfifteen}
\ee

\noi I note that the $\Gamma(K\to \mu\bar\nu(\gamma))$ rate in eq.~(\ref{eqfourteen}) comes from a PDG fit \cite{hagi} rather than a single measurement. One might, therefore, want to expand the error by a factor of 2 or so to be more conservative. However, it would not affect the results presented below in any significant way. It does point out the general need for more precise dedicated measurements of kaon properties. 

Comparing eqs.~(\ref{eqthirteen}) and (\ref{eqfifteen}) leads to the master relation

\be
\frac{|V_{us}|^2 f^2_K}{|V_{ud}|^2 f^2_\pi} = 0.07602 (23) (27) \label{eqsixteen}
\ee

\noi where the errors correspond respectively to the experimental and structure dependent radiative corrections uncertainties. Finally, using the lattice result for $f_K/f_\pi$ in eq.~(\ref{eqthree}) implies

\be
\frac{|V_{us}|^2}{|V_{ud}|^2} = 0.05271 (16) (19) (149) \label{eqseventeen} 
\ee

\noi where the last uncertainty, which clearly dominates, results from combining (in quadrature) the lattice statistical and systematic errors.

Employing the value of $|V_{ud}|$ in eq.~(\ref{eqfive}) results in

\be
|V_{us}| = 0.2236 (1) (3) (4) (30) \label{eqeighteen}
\ee

\noi Combing that value for $|V_{us}|$ with $|V_{ud}|$ from super-allowed nuclear beta decays in eq.~(\ref{eqfive}) gives

\be
|V_{ud}|^2 +|V_{us}|^2 = 0.9987 (17) \label{eqnineteen}
\ee

\noi which is consistent with CKM unitarity expectations. Alternatively, one can assume the unitarity relationship $|V_{ud}|^2 = 1- |V_{us}|^2$ in eq.~(\ref{eqseventeen}) and find

\be
|V_{us}| = 0.2238 (3) (4) (30) \label{eqtwenty}
\ee

\noi independent of the nuclear physics input regarding $|V_{ud}|$. Of course, the proximity of eqs.~(\ref{eqeighteen}) and (\ref{eqtwenty}) is primarily a restatement of the unitarity confirmation in eq.~(\ref{eqnineteen}).

In table~\ref{tabone}, I compare the values of $|V_{us}|$ obtained above with determinations from $K_{e3}$ \cite{hagi,sher} and Hyperon \cite{cabib} decays as well as the indirect determination \cite{renk} from $|V_{ud}|$ assuming CKM unitarity.

\begin{table}
\caption{\label{tabone} Values of $|V_{us}|$ obtained using different approaches.}
\begin{ruledtabular}
\begin{tabular}{ll}
$|V_{us}|$  & Input Employed \\
\hline
0.2236(30) & $f_K/f_\pi + |V_{ud}|$ eq.~(\ref{eqeighteen}) \\
0.2238(30) & $f_K/f_\pi +{}$CKM Unitarity eq.~(\ref{eqtwenty}) \\
0.2196(26) & 2002 PDG $K_{e3}$ Average \protect\cite{hagi} \\
0.2272(30) & E865 $K_{e3}$ \protect\cite{sher} \\
0.2250(27) & Hyperon Decays \protect\cite{cabib} \\
0.2265(22) & $|V_{ud}|+{}$CKM Unitarity \protect\cite{renk} \\
\end{tabular}
\end{ruledtabular}
\end{table}

The lattice based results are consistent with all the other determinations of $|V_{us}|$. In fact, they fall in the middle of the $K_{e3}$ extremes which are individually in disagreement with one another. So, they have not resolved the discrepancy. However, the spread in table~\ref{tabone} values may be indicating that the lattice based result $|V_{us}|\simeq 0.2237$ may be pretty much right on the mark.

In addition to determining $|V_{us}|$, one can use the lattice based value of $f_K/f_\pi$ in conjunction with $|V_{ud}|$ from nuclear beta decay (and indirectly $G_\mu$ from muon decay) to search for or constrain new physics effects via the unitarity expectation $|V_{ud}|^2 +|V_{us}|^2 +|V_{ub}|^2 =1$. A deviation would be indicative of an unaccounted for effect. For example, charged Higgs scalars \cite{chan} or leptoquarks could contribute to the $\pi_{\mu2}$ or $K_{\mu2}$ amplitudes. Alternatively, $Z^\prime$  bosons \cite{marcsix} or exotic muon decay rates \cite{marcseven} could influence the value of $|V_{ud}|$ extracted from $\beta$-decays.

Consider the case of 2 Higgs doublet models \cite{chan} with $\tan\beta=v_2/v_1$ ( the ratio of Higgs vacuum expectations). Its effect on $|V_{ud}|$ would be negligible but its presence in $\pi_{\ell2}$ and $K_{\ell2}$ decay rates (particularly the latter) would imply that the lattice approach advocated here is actually determining $|V_{us}| (1-(m^2_K -m^2_\pi) \tan^2\beta/m^2_{H^+})$. Hence, one would expect a small deviation from CKM unitarity in eq.~(\ref{eqnineteen}) if that scenario were in fact correct. The central value given there is slightly below 1 but consistent with unitarity within the error. Hence, one gets the somewhat diluted constraint

\be
m_{H^+} \gsim  2.5 \tan\beta {\rm ~GeV}\quad (95\% {\rm ~CL}) \label{eqtwentyone}
\ee

\noi which is interesting only for large $\tan \beta \gsim 40$. The unitarity constraint can be applied to leptoquarks, $Z^\prime$ bosons \cite{marcsix}, exotic muon decay rates (eg.\ $\mu^-\to e^-\nu_e \nu_\mu$) \cite{marcseven} etc. To make such constraints really prohibitive the uncertainties in $|V_{ud}|$ and $|V_{us}|$ would both have to be improved by about a factor of 4, a difficult but very well motivated goal. Alternatively, two independent improved determinations of $|V_{us}|$ could be compared.

In summary, the current lattice determination of $f_K/f_\pi$ provides a precise value for $|V_{us}|$ which is already competitive with other more traditional approaches (illustrated in table~\ref{tabone}). It can be used to test CKM unitarity and probe for new physics effects. Perhaps the most interesting aspect of this new approach to $|V_{us}|$ determination is its potential for further improvement. Uncertainties from experiment $(K_{\mu2})$ and structure dependent radiative corrections together constitute at this time an essentially negligible $\pm0.2\%$ error in $|V_{us}|$. So, the lattice error on $f_K/f_\pi$ is dominant and its reduction should occur as lattice calculations become more refined. If the combined error on $f_K/f_\pi$ can be reduced by a factor of 2--4 it should resolve issues regarding CKM unitarity and $K_{e3}$ decay rate discrepencies. Such a reduction may be possible with increased computer power and more sophisticated approaches to chiral symmetry. That potential payoff presents a special opportunity for lattice gauge theory computations to prove their worth.  It should be vigorously pursued.


\end{document}